\documentclass[final]{cvpr}

\usepackage{times}
\usepackage{epsfig}
\usepackage{graphicx}
\usepackage{amsmath}
\usepackage{amssymb}
\usepackage[ruled]{algorithm2e}
\usepackage{multirow}

\usepackage[pagebackref=true,breaklinks=true,colorlinks,bookmarks=false]{hyperref}

\def\cvprPaperID{2067} % *** Enter the CVPR Paper ID here
\def\confYear{CVPR 2021}

\begin{document}

%%%%%%%%% TITLE
\title{Multi-institutional Collaborations for Improving Deep Learning-based Magnetic Resonance Image Reconstruction Using Federated Learning}

\author{Pengfei Guo \qquad Puyang Wang \qquad Jinyuan Zhou \qquad Shanshan Jiang \qquad Vishal M. Patel\\
Johns Hopkins University\\
{\tt\small \{pguo4,pwang47\}@jhu.edu, \{jzhou2,sjiang21\}@jhmi.edu, vpatel36@jhu.edu}	
}

\maketitle

%%%%%%%%% ABSTRACT
\begin{abstract}
	Fast and accurate reconstruction of magnetic resonance (MR) images from under-sampled data is important in many clinical applications. In recent years, deep learning-based methods have been shown to produce superior performance on MR image reconstruction. However, these methods require large amounts of data which is difficult to collect and share due to the high cost of acquisition and medical data privacy regulations.  In order to overcome this challenge, we propose a federated learning (FL) based solution in which we take advantage of the MR data available at different institutions while preserving patients' privacy. However, the generalizability of models trained with the FL setting can still be suboptimal due to domain shift, which results from the data collected at multiple institutions with different sensors, disease types, and acquisition protocols, etc. With the motivation of circumventing this challenge, we propose a cross-site modeling for MR image reconstruction in which the learned intermediate latent features among different source sites are aligned with the distribution of the latent features at the target site. Extensive experiments are conducted to provide various insights about FL for MR image reconstruction. Experimental results demonstrate that the proposed framework is a promising direction to utilize multi-institutional data without compromising patients' privacy for achieving improved MR image reconstruction. Our code will be available at https://github.com/guopengf/FL-MRCM. 
\end{abstract}

%%%%%%%%% BODY TEXT
\section{Introduction}\label{sec1}
\begin{figure}[]
	\centering
	\includegraphics[width=\columnwidth]{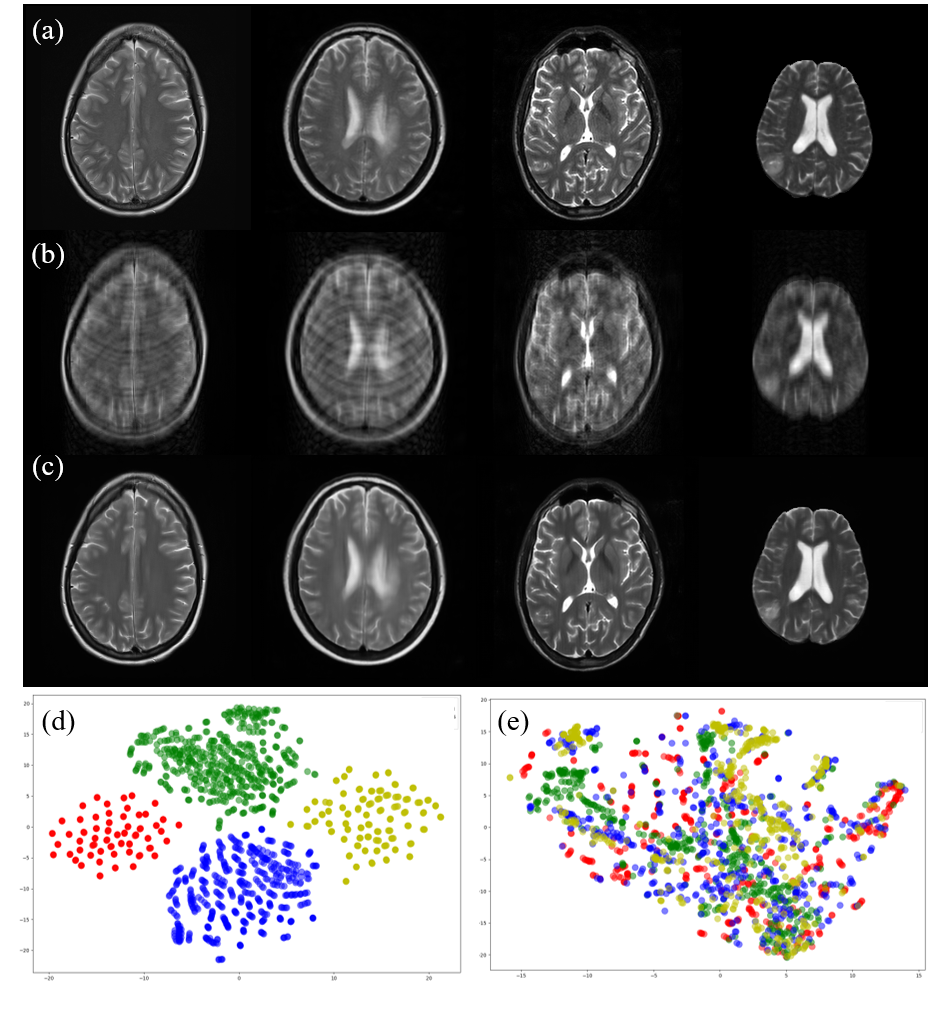}
	\caption{\emph{Top row}: (a) ground truth, (b) zero-filled images, and (c) reconstructed images from the fastMRI~\cite{au45}, HPKS~\cite{au48}, IXI~\cite{au46}, and BraTS~\cite{au47} datasets from left to right, respectively. \emph{Bottom row}: t-SNE plots. The distribution of (d) latent features without cross-site modeling, and (e) latent features corresponding to the proposed cross-site modeling. In each plot, green, blue, yellow, and red dots represent data from fastMRI~\cite{au45}, HPKS~\cite{au48}, IXI~\cite{au46}, and BraTS~\cite{au47} datasets, respectively. \label{fig1}
	}
\end{figure}

Magnetic resonance imaging (MRI) is one of the most widely used imaging techniques in clinical applications.  It is non-invasive and can be customized with different pulse  sequences  to capture different  kinds  of  tissues.  For instance, fat tissues are bright in $T_1$-weighted images, which can clearly show gray and white matter tissues in the brain. The radiofrequency pulse sequences used to make $T_2$-weighted images can delineate fluid from cortical tissue~\cite{au1}. However, to increase the signal-to-noise ratio
(SNR), clinical scanning usually involves the usage of multiple saturation frequencies and repeating acquisitions, which results in relatively long scan time. Various compressed sensing (CS) based methods have been proposed in the literature for accelerating the MRI sampling process by undersampling in the $k$-space during acquisition~\cite{au3, au4, au5}. In recent years, data driven deep learning-based methods have been shown to produce superior performance on MR image reconstruction from partial $k$-space observations~\cite{au6,au7,au8}.   

However, deep networks usually require large amounts of diversity-rich paired data which can be labor-intensive and prohibitively expensive to collect. In addition, one has to deal with patient privacy issues when storing them, making it difficult to share data with other institutions.  Although deidentification~\cite{au15} might provide a solution, building a large scale centralized dataset at a particular institution is still  a challenging task.

The recently introduced federated learning (FL) framework~\cite{au16,au17,au18} addresses this issue by allowing collaborative and decentralized training of deep learning-based methods.
In particular, there is a server that periodically communicates with each institution to aggregate a global model and then shares it with all institutions. Each institution utilizes and stores its own private data.
It is worth noting that instead of directly transferring data for training, the communication in FL algorithms only involves model parameters or update gradients, which resolves the privacy concerns.
Hence, FL methods intrinsically facilitate multi-institutional collaborations between data centers (e.g., hospitals in the context of medical images).  

However, the generalizability of models trained with the FL setting can still be suboptimal due to domain shift, which results from the data collected at multiple institutions with different sensors, disease types, and acquisition protocols, etc. This can be clearly seen from Fig.~\ref{fig1} where we show fully-sampled (Fig.~\ref{fig1}(a)) and under-sampled (Fig.~\ref{fig1}(b)) images from four different datasets. In Fig.~\ref{fig1}(d), we visualize latent features corresponding to images from these datasets using t-Distributed Stochastic Neighbor Embedding (t-SNE) plot~\cite{au20}. As can be seen from Fig.~\ref{fig1}(d), features from a particular dataset are grouped together in a cluster indicating that each dataset has its own bias. As a result, we can see four different clusters of latent features. In order to make use of these datasets in the FL framework, one needs to align these features and remove the domain shift among the datasets. To circumvent this challenge, we propose a cross-site model for MR image reconstruction in which the learned intermediate latent features among different source sites are aligned with the distribution of the latent features at the target site. Specifically, the proposed method involves two optimization steps. In the first step, local reconstruction networks are trained on private data. In the second step, the intermediate latent features of the target domain data are transferred to other local source entities. An adversarial domain identifier is then trained to align the latent space distribution between the source domain and the target domain. Hence, minimizing the loss of adversarial domain identifier results in the reconstruction network weights being automatically adapted to the target domain. Fig.~\ref{fig1}(e) and (c) show the distribution of aligned features and the corresponding reconstructed images in four datasets. The proposed cross-site modeling allows us to leverage datasets from various institutions for obtaining improved reconstructions.  
\begin{figure*}[t!]
	\centering
	\includegraphics[width=\textwidth]{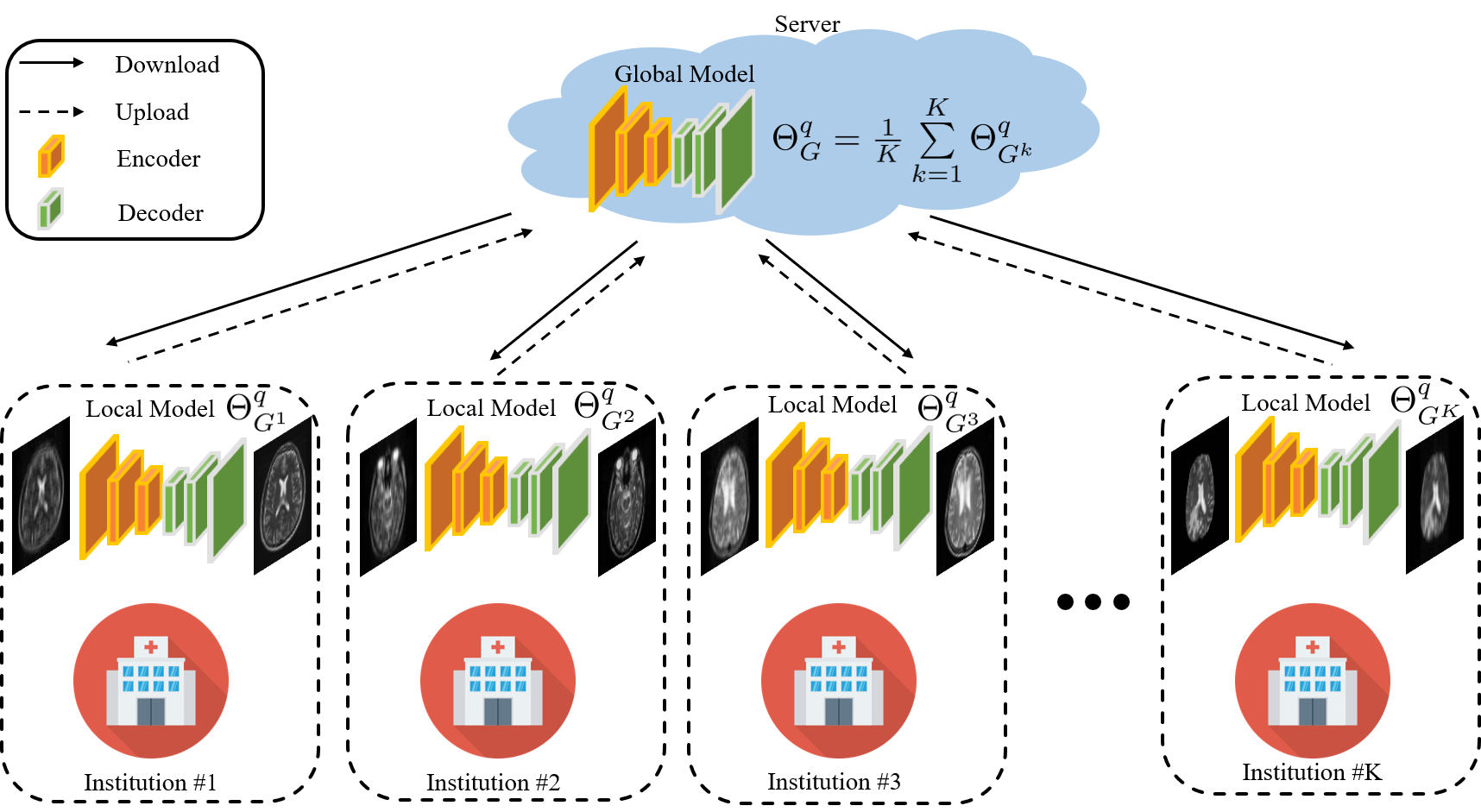}
	\caption{An overview of the proposed FL-MR framework. Through several rounds of communication between data centers and server, the collaboratively trained global model parameterized by $\Theta_G^q$ can be obtained in a data privacy-preserving manner.}\label{fig2}
\end{figure*}

 To summarize, this paper makes the following contributions:
\begin{itemize}
	\item A method called Federated Learning-based Magnetic Resonance Imaging Reconstruction (FL-MR) is proposed which enables multi-institutional collaborations for MRI reconstruction  in a privacy-preserving manner.
	\item To address the domain shift issue among different sites, FL-MR with Cross-site Modeling (FL-MRCM) is proposed to align the latent space distribution between the source domain and the target domain.
	\item Extensive experiments are conducted to provide various insights about FL for MR image reconstruction.  
\end{itemize}
%The rest of the paper is structured as follows. We first show related work on MR image reconstruction and federated learning in Section~\ref{sec2}. We then show the details of our approach in Section~\ref{sec3}. Extensive experiments on our proposed method is shown in Section~\ref{sec4}. We finally conclude our paper in Section~\ref{sec5}.

%------------------------------------------------------------------------

\section{Related Work}\label{sec2}
%\textbf{MR image reconstruction.} 
Reconstruction of MR images from under-sampled $k$-space data is an ill-posed inverse problem.  In order to obtain a regularized solution, some priors are often used.  CS-based methods make use of sparsity priors for recovering the image \cite{au4,au5} from partial $k$-space observations.  In recent years, deep learning-based methods have been shown to produce superior performance on MR image reconstruction \cite{au29,au30,au31,au32,au33}. Some deep learning-based methods approach the problem by directly learning a mapping from the under-sampled data to the fully-sampled data in the image domain~\cite{au7,au8,au34,au35}. Methods that learn a mapping in the  $k$-space domain have also been proposed in the literature~\cite{au36,au37}. 

Federated learning is a decentralized learning framework which allows multiple institutions to collaboratively learn a shared machine learning model without sharing their local training data ~\cite{au24,au16,au23}. The FL training process consists of the following steps: (1) All institutions locally compute gradients and send locally trained network parameters to the server. (2) The server performs aggregation over the uploaded parameters from $K$ institutions. (3) The server  broadcasts the aggregated parameters to $K$ institutions. (4) All institutions update their respective models with aggregated parameters and test the performance of the updated models. The institutions collaboratively learn a machine learning model with the help of a central cloud server ~\cite{au25}.  After a sufficient number of local training and update exchanges between the institutions and the server, a global optimal learned model can be obtained. 

McMahan \emph{et al.}~\cite{au16} proposed FedAvg, which learns a global model by averaging model parameters from local entities. FedAvg~\cite{au16} is one of the most commonly used frameworks for
FL. FedProx \cite{au26} and Agnostic Federated
Learning (AFL)~\cite{au27} are extensions of FedAvg which attempt to address the learning bias issue of the global models for local entities.  Recently, Sheller \emph{et al.}~\cite{au27} and Li \emph{et al.}~\cite{au38} proposed medical image segmentation models based on the FL framework. Peng \emph{et al.}~\cite{au21} proposed the federated adversarial alignment to mitigate the domain shift problem in image classification. In \cite{au22}, Li \emph{et al.} formulated a privacy-preserving pipeline for
multi-institutional functional MRI classification and investigated different aspects of the communication frequency in federated models and privacy-preserving mechanisms. Although these methods~\cite{au21,au22} achieved promising results to overcome domain shift in classification, due to the differences in network architectures, one cannot directly utilize them for MR image reconstruction. It is worth noting that the multi-institutional collaborative approach based on FL for MR image reconstruction has not been well studied in the literature.

%\textbf{Unsupervised domain adaptation.}

%-------------------------------------------------------------------------

\section{Methodology}\label{sec3}
Similar to ~\cite{au7,au8,au34,au35}, the proposed method addresses the MR image reconstruction problem by directly learning a mapping from the under-sampled data to the fully-sampled data in the image domain. The MR image reconstruction process can be formulated as follows
\begin{equation} \label{eq:1}
\begin{aligned}
&x = F^{-1}(F_{d}y + \epsilon),\\
&x, y\in \mathbb{C}^{N},
\end{aligned}
\end{equation}
where $x$ denotes the observed under-sampled image, $y$ is the fully-sampled image, and $\epsilon$ denotes noise. Here, $F$ and $F^{-1}$ denote the Fourier transform matrix and its inverse, respectively. $F_d$ represents the undersampling Fourier encoding matrix that is defined as the multiplication of the Fourier transform matrix $F$ with a binary undersampling mask matrix. The acceleration factor (AF) controls the ratio of the amount of k-space data required for a fully-sampled image to the amount collected in an accelerated acquisition. The goal is to estimate $y$ from the observed under-sampled image $x$.

\subsection{FL-based MRI Reconstruction}
The proposed FL-MR framework is presented in
Fig.~\ref{fig2} and Algorithm~\ref{al:1}. Let $\mathcal{D}^1,\mathcal{D}^2, \dots, \mathcal{D}^K$ denote the MR image reconstruction datasets from $K$ different institutions. Each local dataset $\mathcal{D}^k$ contains pairs of under-sampled and fully-sampled images. At each institution, a local model is trained  using its own data by iteratively minimizing the following loss
\begin{equation} \label{eq:2}
\begin{aligned}
\mathcal{L}_{\text{recon}} &= \sum\limits_{(x,y)\sim \mathcal{D}^k} \|G^k(x)-y\|_1,\\
\end{aligned}
\end{equation}
where $G^k$ corresponds to the local model at site $k$ and is parameterized by $\Theta_{G^k}$. $G^k(x)$ corresponds to the reconstructed image $\hat{y}$. After optimization with several local epochs (i.e. $P$ epochs) via
\begin{equation}\Theta_{G^k}^{(p+1)} \leftarrow \Theta_{G^k}^{(p)}-\gamma\nabla\mathcal{L}_{\text{recon}},
\end{equation}
each institution can obtain the trained FL-MR reconstruction model with the updated model parameters. Since each institution has its own data which may be collected by a particular sensor, disease type, and acquisition protocol, each $\mathcal{D}^k$ has a certain characteristic. Thus, when a local model is trained using its own data, it introduces a bias and does not generalize well to MR images from another institutions (see Fig.~\ref{fig1}). One way to overcome this issue would be to train the network on a diverse multi-domain dataset by combining data from $K$ institutions as $\mathcal{D}=\{\mathcal{D}^1 \cup \mathcal{D}^2 \cup \dots \cup \mathcal{D}^K \}$ \cite{au9,au10,au49,au39,au40}.  However, as discussed earlier, due to privacy concerns, this solution is not feasible and impedes multi-institutional collaborations in practice.

\begin{algorithm}[t!]\label{al:1}
	\SetAlgoLined
	%\justify
	\textbf{Input:} 
	$\mathcal{D} =\{\mathcal{D}^1, \mathcal{D}^2, \dots, \mathcal{D}^K \}$, datasets from K institution; P, the number of local epoches; Q, the number of global epoches; $\gamma$, learning rate; $G^1, G^2, \dots, G^K$, local models parameterized by $\Theta_{G^1},\Theta_{G^2}, \dots, \Theta_{G^K}$; $G$, the global model parameterized by $\Theta_{G}$.\\
	\textbf{Output:} well-trained global model $G$\\
	$\triangleright$ parameters initialization\;
	\For{q = 1 to Q}{
		\For{k = 1 to K \textbf{in parallel}} {
			%$\Theta_{G^k}^{q,(0)} \leftarrow \Theta_{G}^{q} $ \\
			$\triangleright$ \text{deploy weights to local model}\;
			\For{p = 1 to P}{
				
				$\triangleright$ compute reconstruction loss $\mathcal{L}_{\text{recon}}$ with Eq.~\ref{eq:2} and update parameters $\Theta_{G^k}$\;
			}
			%$ \Theta_{G^k}^{q} \leftarrow \Theta_{G^k}^{q,(P)}$	
			$\triangleright$ \text{upload weights to server};
		}
		%$\Theta_{G}^q = \frac{1}{K}\sum\limits_{k=1}^{K}\Theta_{G^k}^q$\\	
		$\triangleright$ \text{update global
			model with Eq.~\ref{eq:3}};
	}
	\Return{$\Theta_{G}^Q$}
	\caption{FL-based MRI Reconstruction}
\end{algorithm}

\begin{figure}[t!]
	\centering
	\includegraphics[width=\columnwidth]{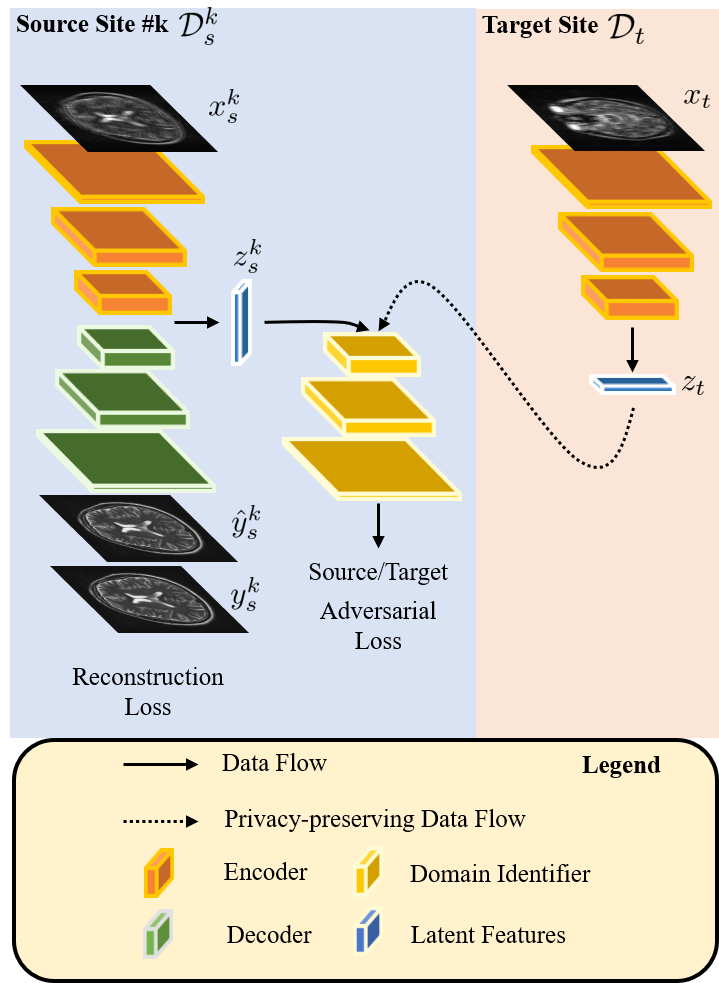}
	\caption{An overview of the proposed FL-MR framework with cross-site modeling in a source site.}\label{fig3}
\end{figure}

To tackle this limitation and allow various sites to collaboratively train a MR image reconstruction model, we propose
the FL-MR framework based on FedAVG~\cite{au16}. Without accessing private data in each site, the proposed FL-MR method leverages a central server to utilize the information from other institutions by aggregating  local model updates. The central server performs the aggregation of model updates by averaging the updated parameters from all local models as follows
\begin{equation} \label{eq:3}
\begin{aligned}
\Theta_{G}^q = \frac{1}{K}\sum\limits_{k=1}^{K}\Theta_{G^k}^q,\\
\end{aligned}
\end{equation}
where $q$ represents the $q$-th global epoch. After $Q$ rounds of communication between local sites and central server, the trained global model parameterized by $\Theta_{G}^Q$, can leverage multi-domain information without directly accessing the private data
in each institution.

\subsection{FL-MR with Cross-site Modeling}
Domain shift among datasets inevitably degrades the performance of machine learning models~\cite{au41}. Existing works~\cite{au42,au43} achieve superior performance by leveraging adversarial training. However, such methods require direct access to the source and target data, which is not allowed in FL-MR. Since we have multiple source domains and the data are stored in local institutions, training a single model that has access to source domains and target domain simultaneously is not feasible. Inspired by federated adversarial alignment~\cite{au21} in classification tasks, we propose FL-MR with Cross-site Modeling (FL-MRCM) to address the domain shift problem in FL-based MRI reconstruction. As shown in Fig.~\ref{fig3}, for a source site $\mathcal{D}_s^k$, we leverage the encoder part of the reconstruction networks ($E_s^k$) to project input onto the latent space $z^k_s$. Similarly, we can obtain $z_t$ for the target site $\mathcal{D}_t$. For each ($\mathcal{D}_s^k$, $\mathcal{D}_t$) source-target domain pair, we introduce an
adversarial domain identifier $\mathcal{C}^k$ to align the latent space distribution between the source domain and the target domain. $\mathcal{C}^k$ is trained in an adversarial manner. Specifically, we first train $\mathcal{C}^k$ to identify which site the latent features come from.  We then train the encoder part of the reconstruction networks to confuse $\mathcal{C}^k$. It should be noted that $\mathcal{C}^k$ only has access to the output latent features from $E_s^k$ and $E_t$, to maintain data sharing regulations. Given the k-th source site data  $\mathcal{D}_s^k$ and the target site data $\mathcal{D}_t$, the loss function for $\mathcal{C}^k$ can be defined as follows
\begin{equation} \label{eq:4}
\begin{aligned}
\mathcal{L}_{\text{adv}\mathcal{C}^k} = &-\mathbb{E}_{x_s^k\sim\mathcal{D}_s^k}[\log \mathcal{C}^k(z^k_s)]\\
&-\mathbb{E}_{x_t\sim\mathcal{D}_t}[\log (1-\mathcal{C}^k(z_t))],
\end{aligned}
\end{equation}
where $z^k_s=E_s^k(x_s^k)$ and $z_t = E_t(x_t) $. The loss function for encoders can be defined as follows
\begin{equation} \label{eq:5}
\begin{aligned}
\mathcal{L}_{\text{adv}E^k} = &-\mathbb{E}_{x_s^k\sim\mathcal{D}_s^k}[\log \mathcal{C}^k(z^k_s)]
-\mathbb{E}_{x_t\sim\mathcal{D}_t}[\log \mathcal{C}^k(z_t)].
\end{aligned}
\end{equation}
The overall loss function used for training the k-th source site with data $\mathcal{D}_s^k$ consists of the reconstruction and adversarial losses. It is defined as follows
\begin{equation} \label{eq:6}
\begin{aligned}
\mathcal{L}_{\text{$\mathcal{D}_s^k$}} = \mathcal{L}_{\text{recon}} + \lambda_{\text{adv}}(\mathcal{L}_{\text{adv}\mathcal{C}^k} + \mathcal{L}_{\text{adv}E^k}),
\end{aligned}
\end{equation}
where $\lambda_{\text{adv}}$ is a constant which controls the contribution of the adversarial loss. The detailed training procedure of FL-MRCM in a source site is presented in Algorithm~\ref{al:2}. In supplementary material, we also provide the schematics of training FL-MRCM in a global view. 
\begin{algorithm}[htp!]\label{al:2}
	\SetAlgoLined
	%\justify
	\textbf{Input:} $\mathcal{D}_s =\{\mathcal{D}^1_s, \mathcal{D}^2_s, \dots, \mathcal{D}^K_s \}$, data from the $K$ source institutions; $\mathcal{D}_t$, data from the target institution; $P$, the number of local epoches; $Q$, the number of global epoches; $\gamma$, learning rate; $\Theta_{G^1_s},..., \Theta_{G^K_s}$, parameters of the local models in the source sites; $\Theta_{\mathcal{C}^1},..., \Theta_{\mathcal{C}^K}$, domain identifiers; $\Theta_{G}$, the global model; $\Theta_{E_t}$, the encoder part of G in the target site.\\
	$\triangleright$ parameters initialization\\
	\For{q = 0 to Q}{
		\For{k = 0 to K \textbf{in parallel}} {
			%$\Theta_{G^k_s}^{q,(0)} \leftarrow \Theta_{G}^{q} $  \\
			$\triangleright$ \text{deploy weights to local model}\\
			\For{p = 0 to P}{
			\textbf{Reconstruction:}\\
			$\triangleright$ compute reconstruction loss $\mathcal{L}_{\text{recon}}$ using Eq.~\ref{eq:2}\\
				\textbf{Cross-site Modeling:}\\
			$\triangleright$ compute adverisal loss $\mathcal{L}_{\text{adv}\mathcal{C}^k}$ and $\mathcal{L}_{\text{adv}E_s^k}$ using Eq.~\ref{eq:4} and Eq.~\ref{eq:5}\\
			$\triangleright$ compute the total loss using Eq.~\ref{eq:6} and update $\Theta_{G_{s}^k}$, $\Theta_{\mathcal{C}^k}$, and $\Theta_{E_t}$
			}
			%$ \Theta_{G^k}^{q} \leftarrow \Theta_{G^k}^{q,(P)}$\\	
			$\triangleright$ \text{upload weights to the central server}
		}
		%$\Theta_{G}^q = \frac{1}{K}\sum\limits_{k=1}^{K}\Theta_{G^k_s}^q$\\
		$\triangleright$ \text{update the global model using Eq.~\ref{eq:3}}\\
	}
	\Return{$\Theta^Q_G$}
	\caption{FL-MR with Cross-site Modeling}
	\vskip-2pt
\end{algorithm}

\subsection{Training and Implementation Details}
We use U-Net~\cite{au44} style encoder-decoder architecture for the reconstruction networks. Details of the network architecture are provided in supplementary material. $\lambda_{\text{adv}}$ is set equal to 1. Acceleration factor (AF) is set equal to 4. The network is trained using the Adam optimizer with the following hyperparameters: constant learning rate of $1\times10^{-4}$ for the first 40 global epochs then $1\times10^{-5}$ for the last global 10 epochs; 50 maximum global epochs; 2 maximum local epochs; batch size of 16. Hyperparameter selection is performed on the IXI validation dataset~\cite{au46}. During training, the cross-sectional images are zero-padded or cropped to the size of 256 $\times$ 256.

%-------------------------------------------------------------------------
\section{Experiments and Results}\label{sec4}
\begin{figure}[t!]
 	\centering
 	\includegraphics[width=\columnwidth]{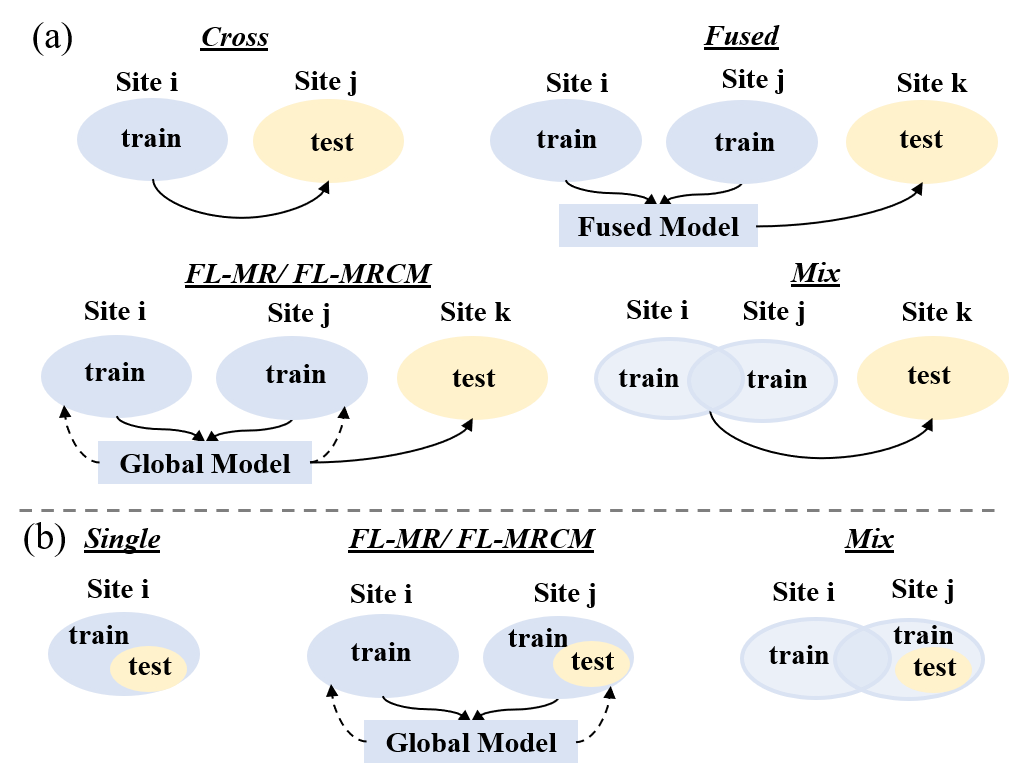}
 	\caption{The schematic of different training strategies in (a) Scenario 1, and (b) Scenario 2. Note that for FL-MRCM, the source sites are the institutions that provide training data and the target site is the institution that provides testing data.}\label{fig4}
\end{figure}

In this section, we present the details of the datasets and
various experiments conducted to demonstrate the effectiveness of the proposed framework. Specifically, we conduct experiments under two scenarios.  Fig.~\ref{fig4} gives an overview of different training and evaluation strategies involved in the two scenarios. In Scenario 1, we analyze the effectiveness of improving the generalizability of the trained models using the proposed methods and other alternative strategies. Thus, the performance of a trained model is evaluated against a dataset that is not directly observed during training.  In particular, we choose one dataset at a time to emulate the role of the user institution and consider data from other sites for training.  This scenario is common in clinical practice. MRI scanners are usually equipped with accelerated acquisition techniques, so the user institution might not have access to fully-sampled data for training. In Scenario 2, we evaluate the proposed method by training it on the data from all available institutions to demonstrate the benefits of collaboration under the setting of federated learning. Rather than assuming that user institution does not have access to fully-sampled data, the training data split of user institution is also involved as a part of collaborations.

\begin{table*}[h!]
	\scriptsize
	\caption{ Quantitative comparison with models trained by different strategies in Scenario 1.}\label{tb1}
	\centering
	\begin{tabular}{c|c|c|c|c|c|c|c|c|c|c}
		\hline
		\multirow{3}{*}{Methods} & \multicolumn{2}{c|}{Data Centers (Institutions)} & \multicolumn{4}{c|}{$T_1$-weighted}                                                                                                     & \multicolumn{4}{c}{$T_2$-weighted}                                                                                                     \\ \cline{2-11} 
		& \multirow{2}{*}{Train}  & \multirow{2}{*}{Test}  & \multirow{2}{*}{SSIM}            & \multirow{2}{*}{PSNR}           & \multicolumn{2}{c|}{Average}                                       & \multirow{2}{*}{SSIM}            & \multirow{2}{*}{PSNR}           & \multicolumn{2}{c}{Average}                                       \\ \cline{6-7} \cline{10-11}
		&                         &                        &                                  &                                 & SSIM                             & PSNR                            &                                  &                                 & SSIM                             & PSNR                            \\ \hline
		\multirow{12}{*}{Cross}  & F                       & B                      & 0.9016                           & 34.65                           & \multirow{12}{*}{0.7907}         & \multirow{12}{*}{30.02}         & 0.9003                           & 33.09                           & \multirow{12}{*}{0.8296}         & \multirow{12}{*}{29.51}         \\
		& H                       & B                      & 0.6670                           & 29.12                           &                                  &                                 & 0.8222                           & 31.06                           &                                  &                                 \\
		& I                       & B                      & 0.8795                           & 33.76                           &                                  &                                 & 0.8610                           & 31.36                           &                                  &                                 \\ \cline{2-5} \cline{8-9}
		& B                       & F                      & 0.7694                           & 28.61                           &                                  &                                 & 0.7851                           & 27.63                           &                                  &                                 \\
		& H                       & F                      & 0.8571                           & 31.82                           &                                  &                                 & 0.8682                           & 29.04                           &                                  &                                 \\
		& I                       & F                      & 0.8417                           & 31.18                           &                                  &                                 & 0.8921                           & 30.08                           &                                  &                                 \\ \cline{2-5} \cline{8-9}
		& B                       & H                      & 0.5188                           & 25.07                           &                                  &                                 & 0.5898                           & 26.28                           &                                  &                                 \\
		& F                       & H                      & 0.8402                           & 28.52                           &                                  &                                 & 0.8842                           & 30.09                           &                                  &                                 \\
		& I                       & H                      & 0.6281                           & 27.09                           &                                  &                                 & 0.8583                           & 29.45                           &                                  &                                 \\ \cline{2-5} \cline{8-9}
		& B                       & I                      & 0.8785                           & 30.10                           &                                  &                                 & 0.7423                           & 27.75                           &                                  &                                 \\
		& F                       & I                      & 0.9102                           & 31.16                           &                                  &                                 & 0.8917                           & 29.57                           &                                  &                                 \\
		& H                       & I                      & 0.7968                           & 29.16                           &                                  &                                 & 0.8598                           & 28.74                           &                                  &                                 \\ \hline
		\multirow{4}{*}{Fused}   & F, H, I                 & B                      & 0.8672                           & 33.98                           & \multirow{4}{*}{0.8223}          & \multirow{4}{*}{31.27}          & 0.8696                           & 32.73                           & \multirow{4}{*}{0.8264}          & \multirow{4}{*}{30.17}          \\
		& B, H, I                 & F                      & 0.8557                           & 32.03                           &                                  &                                 & 0.8524                           & 29.19                           &                                  &                                 \\
		& B, F, I                 & H                      & 0.6615                           & 27.87                           &                                  &                                 & 0.7394                           & 29.28                           &                                  &                                 \\
		& B, F, H                 & I                      & 0.9047                           & 31.22                           &                                  &                                 & 0.8441                           & 29.47                           &                                  &                                 \\ \hline
		\multirow{4}{*}{FL-MR}   & F, H, I                 & B                      & 0.9452                           & 35.59                           & \multirow{4}{*}{0.8976}          & \multirow{4}{*}{32.09}          & 0.916                            & 33.76                           & \multirow{4}{*}{0.8997}          & \multirow{4}{*}{31.49}          \\
		& B, H, I                 & F                      & 0.9099                           & 33.15                           &                                  &                                 & 0.8991                           & 30.86                           &                                  &                                 \\
		& B, F, I                 & H                      & 0.8249                           & 28.49                           &                                  &                                 & 0.8874                           & 31.02                           &                                  &                                 \\
		& B, F, H                 & I                      & 0.9103                           & 31.11                           &                                  &                                 & 0.8962                           & 30.32                           &                                  &                                 \\ \hline
		\multirow{4}{*}{FL-MRCM} & F, H, I                 & B                      & \textbf{0.9504} & \textbf{35.93} & \multirow{4}{*}{\textbf{0.9108}} & \multirow{4}{*}{\textbf{32.51}} & \textbf{0.9275} & \textbf{33.96} & \multirow{4}{*}{\textbf{0.9113}} & \multirow{4}{*}{\textbf{31.77}} \\
		& B, H, I                 & F                      & \textbf{0.9149} & \textbf{33.31} &                                  &                                 & \textbf{0.9139} & \textbf{31.31} &                                  &                                 \\
		& B, F, I                 & H                      & \textbf{0.8581} & \textbf{29.24} &                                  &                                 & \textbf{0.8978} & \textbf{31.35} &                                  &                                 \\
		& B, F, H                 & I                      & \textbf{0.9197} & \textbf{31.54} &                                  &                                 & \textbf{0.9058} & \textbf{30.47} &                                  &                                 \\ \hline\hline
		& F, H, I                 & B                      & 0.9589                           & 36.68                           & \multirow{4}{*}{0.9182}          & \multirow{4}{*}{32.96}          & 0.9464                           & 34.58                           & \multirow{4}{*}{0.9260}          & \multirow{4}{*}{32.44}          \\
		Mix                      & B, H, I                 & F                      & 0.9222                           & 33.79                           &                                  &                                 & 0.9239                           & 31.89                           &                                  &                                 \\
		(Upper Bound)            & B, F, I                 & H                      & 0.8630                           & 29.19                           &                                  &                                 & 0.9168                           & 32.14                           &                                  &                                 \\
		& B, F, H                 & I                      & 0.9286                           & 32.19                           &                                  &                                 & 0.9169                           & 31.14                           &                                  &                                 \\ \hline
	\end{tabular}
\end{table*}
\subsection{Datasets}
\noindent \textbf{fastMRI}~\cite{au45} (F for short):
$T_1$-weighted images corresponding to
3443 subjects are used for conducting experiments.  In particular, data from 2583 subjects are used
for training and remaining data from 860 subjects are used
for testing. In addition, $T_2$-weighted images from
3832 subjects are also used, where data from 2874 subjects are used
for training and data from 958 subjects are used
for testing. For each subject, approximately 15 axial cross-sectional images that contain brain tissues are provided in this dataset.

\noindent\textbf{HPKS}~\cite{au48} (H for short): This dataset is collected from post-treatment patients with malignant glioma. $T_1$ and  $T_2$-weighted images from 144 subjects are analyzed, where 116 subjects' data are used for training and 28 subjects' data are used for testing. For each subject, 15 axial cross-sectional images that contain brain tissues are provided in this dataset.

\noindent\textbf{IXI}~\cite{au46} (I for short):
$T_1$-weighted images from
581 subjects are used, where 436 subjects' data are used
for training, 55 subjects' data are used
for validation, and 90 subjects' data are used
for testing. $T_2$-weighted images from
578 subjects are also analyzed, where 434 subjects' data are used
for training, 55 subjects' data are used
for validation and the remaining 89 subjects' data are used
for testing. For each subject, there are approximately 150 and 130 axial cross-sectional images that contain brain tissues for $T_1$ and $T_2$-weighted MR sequences, respectively.

\noindent\textbf{BraTS}~\cite{au47} (B for short): $T_1$ and  $T_2$-weighted images from 494 subjects are used, where 369 subjects' data are used
for training and 125 subjects' data are used for testing. For each subject, approximately 120 axial cross-sectional images that contain brain tissues are provided for both MR sequences.

 \begin{figure*}[t!]
	\centering
	\includegraphics[width=\textwidth]{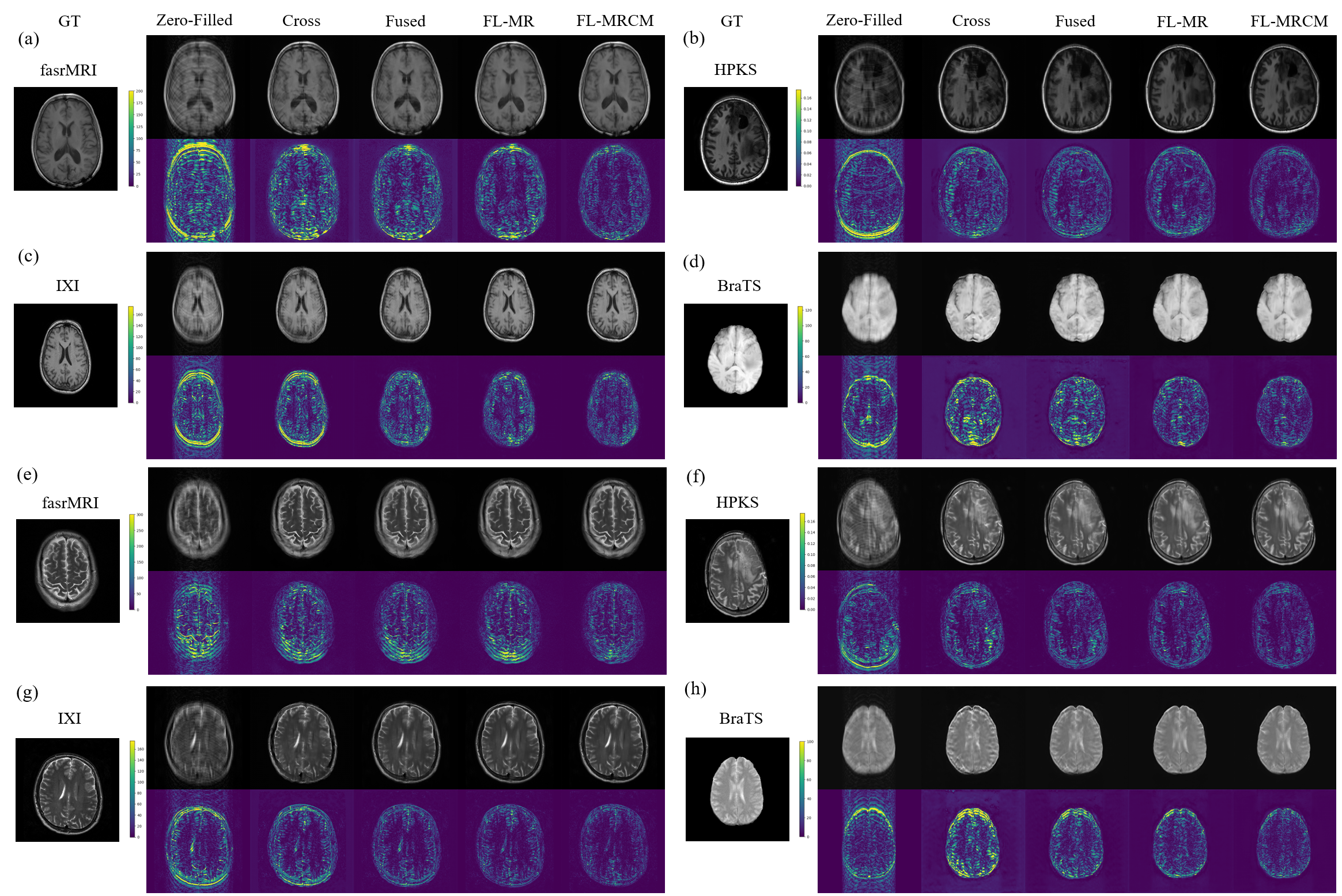}
	\caption{Qualitative results of different methods that correspond to Scenario 1. For results of $T_1$-weighted images on, (a) fastMRI~\cite{au45}, (b) HPKS~\cite{au48}, (c) IXI~\cite{au46}, (d) BraTS~\cite{au47}.  For results of $T_2$-weighted images on, (e) fastMRI~\cite{au45}, (f) HPKS~\cite{au48}, (g) IXI~\cite{au46}, (h) BraTS~\cite{au47}. The second row of each sub-figure shows the absolute image difference between reconstructed images and the ground truth.}\label{fig5}
\end{figure*}
\subsection{Evaluation of the Generalizability}
In the first set of experiments (Scenario 1), we analyze the model's generalizability to data from another site. In Table~\ref{tb1}, we compare the quality of reconstructed images from different methods on four datasets using structural similarity index measure (SSIM) and peak-signal-to-noise ratio (PSNR). We first compare the performance of the proposed framework with models trained with data from a single data center. In this case, we obtain a trained model from one of the institutions and evaluate its performance on another data center in Table~\ref{tb1} under the label \textbf{Cross}. It is also possible to obtain multiple trained
models from several institutions and fuse their outputs, which does not violate privacy regulations.
In this case, we fuse the reconstructed images of the trained
model from various institutions by calculating the average.
The results corresponding to this strategy are shown in Table~\ref{tb1} under the label \textbf{Fused}. In addition, we can obtain a model that is trained with data from all available data centers, which is
denoted by \textbf{Mix} in Table~\ref{tb1}. However, this case compromises subjects' privacy from other institutions, so we treat it as an upper bound. 

As it can be seen from Table~\ref{tb1}, our proposed FL-MR method exhibits better generalization and clearly outperforms other privacy-preserving alternative strategies. FL-MRCM further improves the reconstruction quality in each dataset by mitigating the domain shift. Fig.~\ref{fig5} shows the qualitative performance of different methods on $T_1$ and $T_2$-weighted images from four datasets. It can be observed that the proposed FL-MRCM method yields reconstructed images with remarkable visual similarity to the reference images compared to the other alternatives (see the last column of each sub-figure in Fig.~\ref{fig5}) in four datasets with diverse characteristics. 

\begin{table}[ht!]
	\setlength{\tabcolsep}{1.0pt}
	\scriptsize
	\caption{ Quantitative comparison with models trained by different strategies in Scenario 2.}\label{tb2}
	\centering
	\begin{tabular}{ccc|c|c|c|c|c|c|c|c}
		\hline
		\multicolumn{1}{c|}{\multirow{4}{*}{Methods}} & \multicolumn{2}{c|}{Data Centers}                                        & \multicolumn{4}{c|}{\multirow{2}{*}{$T_1$-weighted}}                                                               & \multicolumn{4}{c}{\multirow{2}{*}{$T_2$-weighted}}                                                               \\
		\multicolumn{1}{c|}{}                         & \multicolumn{2}{c|}{(Institutions)}                                       & \multicolumn{4}{c|}{}                                                                                              & \multicolumn{4}{c}{}                                                                                              \\ \cline{2-11} 
		\multicolumn{1}{c|}{}                         & \multicolumn{1}{c|}{\multirow{2}{*}{Tain}}       & \multirow{2}{*}{Test} & \multirow{2}{*}{SSIM} & \multirow{2}{*}{PSRN} & \multicolumn{2}{c|}{Average}                                       & \multirow{2}{*}{SSIM} & \multirow{2}{*}{PSNR} & \multicolumn{2}{c}{Average}                                       \\ \cline{6-7} \cline{10-11} 
		\multicolumn{1}{c|}{}                         & \multicolumn{1}{c|}{}                            &                       &                       &                       & SSIM                             & PSNR                            &                       &                       & SSIM                             & PSNR                            \\ \hline
		\multicolumn{1}{c|}{\multirow{4}{*}{Single}}  & \multicolumn{1}{c|}{B}                           & B                     & 0.9660                & 37.30                 & \multirow{4}{*}{0.9351}          & \multirow{4}{*}{33.81}          & 0.9558                & 34.90                 & \multirow{4}{*}{0.9278}          & \multirow{4}{*}{32.35}          \\
		\multicolumn{1}{c|}{}                         & \multicolumn{1}{c|}{F}                           & F                     & 0.9494                & 35.45                 &                                  &                                 & 0.9404                & 32.43                 &                                  &                                 \\
		\multicolumn{1}{c|}{}                         & \multicolumn{1}{c|}{H}                           & H                     & 0.8855                & 29.67                 &                                  &                                 & 0.9001                & 31.29                 &                                  &                                 \\
		\multicolumn{1}{c|}{}                         & \multicolumn{1}{c|}{I}                           & I                     & 0.9396                & 32.80                 &                                  &                                 & 0.9151                & 30.79                 &                                  &                                 \\ \hline
		\multicolumn{1}{c|}{\multirow{4}{*}{FL-MR}}   & \multicolumn{1}{c|}{\multirow{4}{*}{B, F, H, I}} & B                     & 0.9662                & 37.37                 & \multirow{4}{*}{0.9294}          & \multirow{4}{*}{33.92}          & 0.9482                & 35.34                 & \multirow{4}{*}{0.9238}          & \multirow{4}{*}{32.64}          \\
		\multicolumn{1}{c|}{}                         & \multicolumn{1}{c|}{}                            & F                     & 0.9404                & 35.25                 &                                  &                                 & 0.9306                & 32.19                 &                                  &                                 \\
		\multicolumn{1}{c|}{}                         & \multicolumn{1}{c|}{}                            & H                     & 0.8732                & 30.03                 &                                  &                                 & 0.9021                & 31.74                 &                                  &                                 \\
		\multicolumn{1}{c|}{}                         & \multicolumn{1}{c|}{}                            & I                     & 0.9379                & 33.03                 &                                  &                                 & 0.9145                & 31.29                 &                                  &                                 \\ \hline
		\multicolumn{1}{c|}{\multirow{4}{*}{FL-MRCM}} & \multicolumn{1}{c|}{\multirow{4}{*}{B, F, H, I}} & B                     & 0.9676                & 37.57                 & \multirow{4}{*}{\textbf{0.9381}} & \multirow{4}{*}{\textbf{34.14}} & 0.9630                & 35.85                 & \multirow{4}{*}{\textbf{0.9373}} & \multirow{4}{*}{\textbf{33.13}} \\
		\multicolumn{1}{c|}{}                         & \multicolumn{1}{c|}{}                            & F                     & 0.9475                & 35.57                 &                                  &                                 & 0.9385                & 32.69                 &                                  &                                 \\
		\multicolumn{1}{c|}{}                         & \multicolumn{1}{c|}{}                            & H                     & 0.8940                & 30.27                 &                                  &                                 & 0.9232                & 32.44                 &                                  &                                 \\
		\multicolumn{1}{c|}{}                         & \multicolumn{1}{c|}{}                            & I                     & 0.9432                & 33.13                 &                                  &                                 & 0.9244                & 31.54                 &                                  &                                 \\ \hline\hline
		\multicolumn{1}{c|}{}                         & \multicolumn{1}{c|}{\multirow{4}{*}{B, F, H, I}} & B                     & 0.9698                & 37.62                 & \multirow{4}{*}{0.9440}          & \multirow{4}{*}{34.35}          & 0.9655                & 35.83                 & \multirow{4}{*}{0.9398}          & \multirow{4}{*}{33.14}          \\
		\multicolumn{1}{c|}{Mix}                      & \multicolumn{1}{c|}{}                            & F                     & 0.9558                & 36.15                 &                                  &                                 & 0.9435                & 32.82                 &                                  &                                 \\
		\multicolumn{1}{c|}{(Upper}                   & \multicolumn{1}{c|}{}                            & H                     & 0.9047                & 30.57                 &                                  &                                 & 0.9236                & 32.47                 &                                  &                                 \\
		\multicolumn{1}{c|}{Bound)}                   & \multicolumn{1}{c|}{}                            & I                     & 0.9454                & 33.08                 &                                  &                                 & 0.9266                & 31.44                 &                                  &                                 \\ \hline
	\end{tabular}
\end{table}

\begin{figure}[h!]
	\includegraphics[width=\columnwidth]{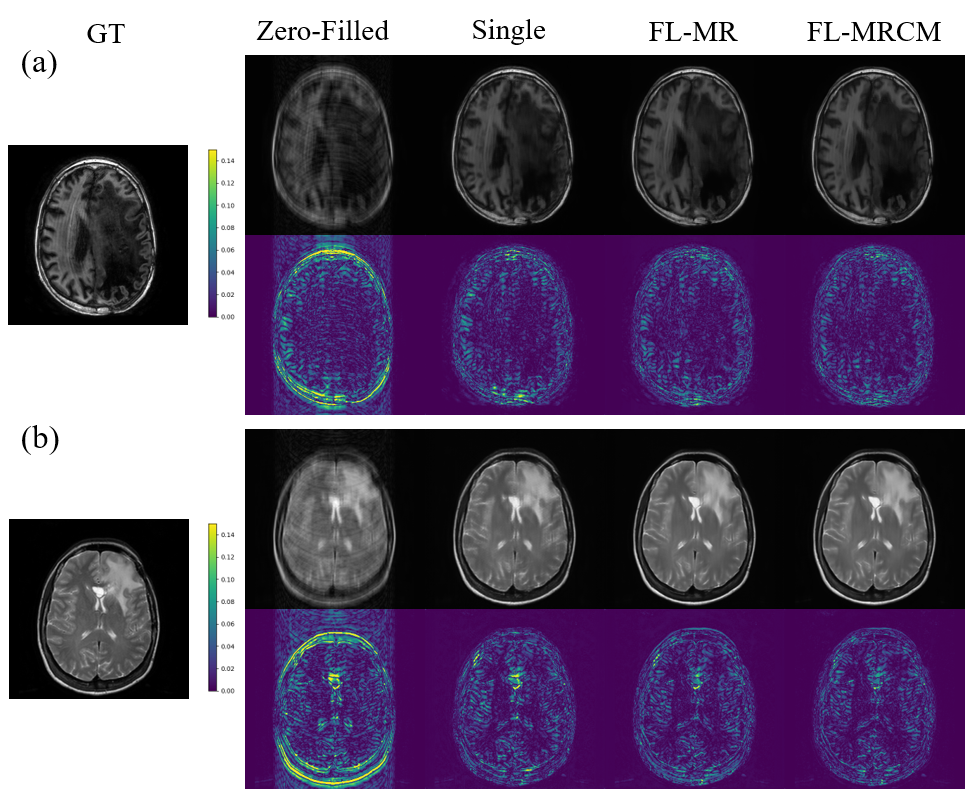}
	\caption{Qualitative results and error maps corresponding to different methods in Scenario 2 on HPKS~\cite{au48}. (a) $T_1$- weighted, and (b) $T_2$-weighted images.}\label{fig6}
\end{figure}

\subsection{Evaluation of FL-based Collaborations}
In the second set of experiments (Scenario 2), we analyze the effectiveness of our method to leverage data from all available institutions in a privacy-preserving manner. 
Since the goal is to evaluate the benefit of multi-institution collaborations, we compare the performance of the proposed framework with models trained with data from a single data center and evaluate on its own testing data, which is
denoted by \textbf{Single} in Table~\ref{tb2}. Similar to Scenario 1, we obtain a model that is directly trained with all available data, which is denoted by \textbf{Mix} in Table~\ref{tb2} and we treat it as an upper bound. It can be seen that the proposed FL-MRCM method outperforms the other methods and reaches the upper bound in term of SSIM and PSNR. It is worth noting that the multi-institution collaborations by the proposed FL-based method exhibits significant improvement on the smaller dataset. Specifically, on the HPKS~\cite{au48}, FL-MRCM improves SSIM from 0.9001 to 0.9232 and PSNR from 31.29 to 32.44 in $T_2$-weighted sequences. As shown in Fig.~\ref{fig6}, the proposed methods have a better ability of suppressing errors around the skull and lesion regions, which is consistent with the quantitative results.

\subsection{Ablation Study}
\begin{figure}[t!]
	\centering
	\includegraphics[width=\columnwidth]{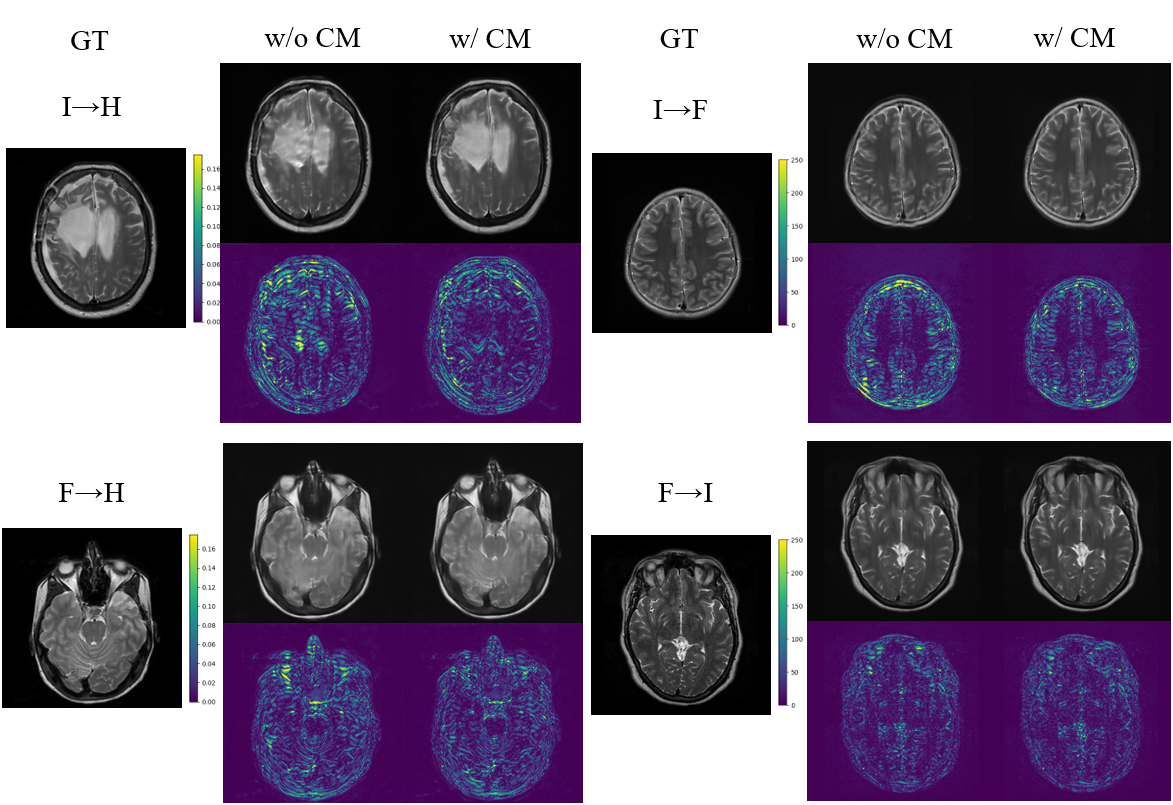}
	\caption{Qualitative comparisons and error maps on the $T_2$-weighted images using cross-site modeling (CM). I$\rightarrow$H represents the results from the model trained on I and tested on H, etc.}\label{fig7}
\end{figure}

The individual contribution of proposed cross-site modeling is demonstrated by a set of experiments (i.e. the comparison between FL-MR and FL-MRCM) in two scenarios under the setting of FL. Furthermore, we conduct a detailed ablation study to analyze the effectiveness of the proposed cross-site modeling without the FL framework. In this case, we obtain a trained model from one of the available sites and evaluate its performance on the data from another institution to observe the gain purely contributed by the cross-site modeling in Table~\ref{tb3}. Sample reconstructed images are shown in Fig.~\ref{fig7}. Experiments with cross-site modeling achieve smaller error. Due to space constraint, a similar ablation study on $T_1$-weighted images, more experimental results, and visualizations are provided in supplementary material.

\begin{table}[t!]
	\setlength{\tabcolsep}{2.9pt}
	\scriptsize
	\caption{ Quantitative ablation study of the proposed cross-site modeling on the $T_2$-weighted images. For experiments with cross-site modeling, the target site is the institution that provides the test data.}\label{tb3}
	\centering
\begin{tabular}{cc|c|c|c|c|c|c|c|c}
	\hline
	\multicolumn{2}{c}{Data Centers}              & \multicolumn{4}{|c|}{w/o Cross-site Modeling}                                                     & \multicolumn{4}{c}{w/ Cross-site Modeling}                                                                        \\ \cline{3-10}
	\multicolumn{2}{c|}{(Institutions)} & \multirow{2}{*}{SSIM} & \multirow{2}{*}{PSNR} & \multicolumn{2}{c|}{Average}                     & \multirow{2}{*}{SSIM} & \multirow{2}{*}{PSNR} & \multicolumn{2}{c}{Average}                                       \\ \cline{1-2} \cline{5-6} \cline{9-10} 
	\multicolumn{1}{c|}{Train} & Test &                       &                       & SSIM                    & PSNR                   &                       &                       & SSIM                             & PSNR                            \\ \hline
	\multicolumn{1}{c|}{B}     & F    & 0.7851                & 27.63                 & \multirow{3}{*}{0.7057} & \multirow{3}{*}{27.22} & 0.7914                & 27.85                 & \multirow{3}{*}{\textbf{0.7525}} & \multirow{3}{*}{\textbf{27.32}} \\
	\multicolumn{1}{c|}{B}     & H    & 0.5898                & 26.28                 &                         &                        & 0.6806                & 26.08                 &                                  &                                 \\
	\multicolumn{1}{c|}{B}     & I    & 0.7423                & 27.75                 &                         &                        & 0.7856                & 28.03                 &                                  &                                 \\ \hline
	\multicolumn{1}{c|}{F}     & B    & 0.9003                & 33.09                 & \multirow{3}{*}{0.8921} & \multirow{3}{*}{30.92} & 0.9139                & 33.84                 & \multirow{3}{*}{\textbf{0.9027}} & \multirow{3}{*}{\textbf{31.58}} \\
	\multicolumn{1}{c|}{F}     & H    & 0.8842                & 30.09                 &                         &                        & 0.8936                & 30.75                 &                                  &                                 \\
	\multicolumn{1}{c|}{F}     & I    & 0.8917                & 29.57                 &                         &                        & 0.9004                & 30.14                 &                                  &                                 \\ \hline
	\multicolumn{1}{c|}{H}     & B    & 0.8222                & 31.06                 & \multirow{3}{*}{0.8501} & \multirow{3}{*}{29.61} & 0.8391                & 31.54                 & \multirow{3}{*}{\textbf{0.8582}} & \multirow{3}{*}{\textbf{30.07}} \\
	\multicolumn{1}{c|}{H}     & F    & 0.8682                & 29.04                 &                         &                        & 0.8646                & 29.36                 &                                  &                                 \\
	\multicolumn{1}{c|}{H}     & I    & 0.8598                & 28.74                 &                         &                        & 0.8709                & 29.31                 &                                  &                                 \\ \hline
	\multicolumn{1}{c|}{I}     & B    & 0.8610                & 31.36                 & \multirow{3}{*}{0.8738} & \multirow{3}{*}{30.30} & 0.8946                & 32.11                 & \multirow{3}{*}{\textbf{0.8949}} & \multirow{3}{*}{\textbf{31.06}} \\
	\multicolumn{1}{c|}{I}     & F    & 0.8921                & 30.08                 &                         &                        & 0.9065                & 30.80                 &                                  &                                 \\
	\multicolumn{1}{c|}{I}     & H    & 0.8583                & 29.45                 &                         &                        & 0.8837                & 30.26                 &                                  &                                 \\ \hline
\end{tabular}
\vskip-5pt
\end{table}
%-------------------------------------------------------------------------
\section{Conclusion}\label{sec5}
We present a FL-based framework to leverage multi-institutional data for the MR image reconstruction task in a privacy-preserving manner. To address the domain shift issue during collaborations, we introduce a cross-site modeling approach that provides the supervision to align the latent space distribution between the source domain and the target domain in each local entity without directly sharing the data. Through extensive experiments on four datasets with diverse characteristics, it is demonstrated that the proposed method is able to achieve better generalization. In addition, we show the benefits of multi-institutional collaborations under the FL-based framework in MR image reconstruction task.
{\small
\bibliographystyle{ieee_fullname}
\bibliography{egbib}
}

\end{document}